\documentclass[conference]{IEEEtran}

\ifCLASSINFOpdf
\else
\fi

\usepackage{amsmath, amssymb, bm, cite, epsfig}
\usepackage{epstopdf}
\usepackage{graphicx}
\usepackage{array}
\usepackage{multirow}
\renewcommand{\arraystretch}{1.5}
\usepackage{amsfonts}
\usepackage{tabularx} 
\usepackage{tabu}

\def\beq{\begin{equation}}
\def\eeq{\end{equation}}
\def\beqa{\begin{eqnarray}}
\def\eeqa{\end{eqnarray}}
\def\beqan{\begin{eqnarray*}}
\def\eeqan{\end{eqnarray*}}

\def\PL{\mathrm{PL}}
\def\dB{\mathrm{dB}}

\def\tm1{t\! - \! 1}
\def\tp1{t\! + \! 1}

\def\PL{\mathrm{PL}}
\def\dB{\mathrm{dB}}
\def\FSPL{\mathrm{FSPL}}
\def\SF{\mathrm{SF}}
\def\ABG{\mathrm{ABG}}
\def\CI{\mathrm{CI}}
\def\FI{\mathrm{FI}}

\usepackage{tikz}
\usetikzlibrary{calc}

\begin{document}
\title{Path Loss, Shadow Fading, and Line-Of-Sight Probability Models for 5G Urban Macro-Cellular Scenarios}

%\author{\IEEEauthorblockN{Michael Shell}
%\IEEEauthorblockA{School of Electrical and\\Computer Engineering\\
%Georgia Institute of Technology\\
%Atlanta, Georgia 30332--0250\\
%Email: http://www.michaelshell.org/contact.html}
%\and
%\IEEEauthorblockN{Homer Simpson}
%\IEEEauthorblockA{Twentieth Century Fox\\

\author{\IEEEauthorblockN{Shu Sun$^{a*}$, Timothy A. Thomas$^b$, Theodore S. Rappaport$^a$, Huan Nguyen$^{c}$, Istv$\acute{a}$n Z. Kov$\acute{a}$cs$^{d}$, and Ignacio Rodriguez$^{c}$}
\IEEEauthorblockA{$^a$NYU WIRELESS and Polytechnic School of Engineering, New York University, Brooklyn, NY, USA 11201\\
%Email: http://www.michaelshell.org/contact.html}
$^b$Nokia, Arlington Heights, IL, USA 60004\\
$^{c}$Aalborg University, Aalborg, Denmark 9220\\
$^{d}$Nokia, Aalborg, Denmark 9220\\
$^*$Corresponding author: ss7152@nyu.edu }}

\maketitle
\begin{tikzpicture}[remember picture, overlay]
\node at ($(current page.north) + (0,-0.25in)$) {S. Sun \textit{et al.}, \rq\rq{}Path Loss, Shadow Fading, and Line-Of-Sight Probability Models for 5G Urban Macro-Cellular Scenarios,\rq\rq{}};
\node at ($(current page.north) + (0,-0.4in)$) {to appear in \textit{2015 IEEE Global Communications Conference Workshop (Globecom Workshop)}, Dec. 2015.};
%\node at ($(current page.north) + (0,-0.55in)$) {\textit{(Globecom Workshop)}, Dec. 2015.};
\end{tikzpicture}

\begin{abstract}
This paper presents key parameters including the line-of-sight (LOS) probability, large-scale path loss, and shadow fading models for the design of future fifth generation (5G) wireless communication systems in urban macro-cellular (UMa) scenarios, using the data obtained from propagation measurements at 38 GHz in Austin, US, and  at 2, 10, 18, and 28 GHz in Aalborg, Denmark. A comparison of different LOS probability models is performed for the Aalborg environment. Alpha-beta-gamma and close-in reference distance path loss models are studied in depth to show their value in channel modeling. Additionally, both single-slope and dual-slope omnidirectional path loss models are investigated to analyze and contrast their root-mean-square (RMS) errors on measured path loss values. While the results show that the dual-slope large-scale path loss model can slightly reduce RMS errors compared to its single-slope counterpart in non-line-of-sight (NLOS) conditions, the improvement is not significant enough to warrant adopting the dual-slope path loss model. Furthermore, the shadow fading magnitude versus distance is explored, showing a slight increasing trend in LOS and a decreasing trend in NLOS based on the Aalborg data, but more measurements are necessary to gain a better knowledge of the UMa channels at centimeter- and millimeter-wave frequency bands. 
\end{abstract}

\IEEEpeerreviewmaketitle
\section{Introduction}
With the rapid growth of personal communication devices such as smart phones and tablets, and with consumers demanding more data access, higher data rates and quality, industry is motivated to develop disruptive technologies and deploy new frequency bands that give rise to the fifth generation (5G) wireless communications. The communication scenarios envisioned for 5G are likely to be similar to those defined in current 4G systems\cite{3GPP:25996,WINNER}, embracing urban micro- (UMi) and urban macro- (UMa) cellular scenarios, indoor hotspot (InH) scenarios, etc. 

Fundamental changes in system and network design will occur in 5G due to emerging revolutionary technologies, potential new spectra such as millimeter-wave (mmWave) frequencies\cite{Rap15}, and novel architectural concepts\cite{And14,Boc14}, thus it is vital to establish reliable channel models to assist engineers in the design. Channel characterization at both mmWave and centimeter-wave (cmWave) bands has been conducted by many prior researchers. The authors in\cite{Soma00,Geng09,Xu02} studied and modeled the UMi and indoor channels at 28 GHz and 60 GHz. Extensive propagation measurements have been carried out recently at 28 GHz, 38 GHz, and 73 GHz in UMi, UMa, and/or indoor scenarios\cite{Rap13:Access,RapGut13,Mac14,Rap15:TCOM,MacCartney15_2}, from which spatial and temporal statistics were extracted in conjunction with the ray-tracing technique. Line-of-sight (LOS) probabilities, directional and omnidirectional path loss models in dense urban environments at 28 GHz and 73 GHz have been investigated in\cite{Mac14:PIMRC,MKS:WCL15}. Two-dimensional (2D) and 3D 28 GHz statistical spatial channel models (SSCMs) have been developed in\cite{Sam14,Sam15:GCWS} that could accurately reproduce wideband power delay profiles (PDPs), angle of departure (AoD), and angle of arrival (AoA) power spectra. 3GPP\cite{3GPP:25996} and WINNER II\cite{WINNER} channel models are the most well-known and widely employed models, containing a variety of communication scenarios including UMi, UMa, indoor office, indoor shopping mall, and so on, and provide important channel parameters such as path loss models, path delays, path powers, and LOS probabilities. However, the 3GPP and WINNER models are only applicable for bands below 6 GHz and hence all of the modeling needs to be revisited for bands above 6 GHz. 

A majority of the previous path loss models are of single-slope, i.e., the model uses one single slope to represent path loss or received power over the entire distance range. While the single slopes are easy to model and have simple mathematical expressions, the root-mean-square (RMS) error between the path loss equation and the local path loss values, often regarded as a measure of shadow fading, can be large for wide ranges of transmitter-receiver (T-R) separation distances, especially in non-line-of-sight (NLOS) environments. This has led to the idea of dual-slope path loss models, which apply different slopes for different regions of T-R separation distances, aimed to reduce the RMS error. Dual-slope path loss models were first proposed and studied in\cite{Bla92,Feu94} for the \textit{close-in} (CI) free space reference distance path loss model in LOS environments, where two double regression approaches were explored with one employing a breakpoint at the first Fresnel zone distance, and the other using a breakpoint determined by the minimum mean square error (MMSE) fitting. The dual-slope model on the basis of the \textit{floating intercept} (FI) path loss model in NLOS environments has been presented in\cite{Soo15:JSTSP}, showing the potential of the dual-slope approach in reducing the RMS error. In addition, geometry-induced shadow fading was derived and modeled as a function of distance in\cite{Soo15:JSTSP} based on the distance-dependency characteristic of shadow fading. 

In this paper, we present propagation measurements conducted in 2011 at 38 GHz in Austin, US\cite{Rap15:TCOM}, and at 2 GHz, 10 GHz, 18 GHz, and 28 GHz in Aalborg, Denmark, in 2015, in UMa environments (where the transmitter height is typically 25 m or so, and the minimum 2D T-R separation distance is 35 m\cite{3GPP_LTE}) (It is suggested that future 3GPP consider 3D distances given the directional nature of future mmWave antennas and sensitivity to pointing angles.) The LOS probability, single-slope multi-frequency \textit{alpha-beta-gamma} (ABG) and CI path loss models, single- and dual-slope path loss models, and distance-dependent shadow fading are studied to gain some insights on large-scale propagation characteristics and to assist in 5G UMa channel modeling. In 5G wireless systems, multiple-input multiple-output (MIMO) systems including beamforming functions have been envisioned as a key component, hence angular statistics of communication channels such as the distributions of AoD, AoA, and angular spread are worth studying, but this is beyond the scope of this paper and can be considered in future work.

\section{Propagation Measurements in UMa Scenarios}\label{Measurements}
In this section, we present two propagation measurement campaigns in outdoor UMa scenarios conducted at the campus of The University of Texas at Austin (UT Austin) in US and Aalborg University (AAU) in Denmark, respectively. 
\subsection{UMa Measurements at UT Austin}
In the summer of 2011, 38 GHz propagation measurements were conducted with four transmitter (TX)  locations chosen on buildings at the UT Austin campus\cite{RapGut13,Rap15:TCOM}, using a spread spectrum sliding correlator channel sounder and directional steerable high-gain horn antennas, with a maximum RF transmit power of 21.2 dBm over an 800 MHz first null-to-null RF bandwidth and a maximum measurable dynamic range of 160 dB, for receiver (RX) locations in the surrounding campus. The measurements used narrowbeam TX antennas (7.8$^{\circ}$ azimuth half-power beamwidth (HPBW)) and narrowbeam (7.8$^{\circ}$ azimuth HPBW) or widebeam (49.4$^{\circ}$ azimuth HPBW) RX antennas. Among the four TX sites, three were with heights of 23 m or 36 m, representing the typical heights of base stations in UMa scenarios. A total of 33 TX-RX location combinations were measured using the narrowbeam RX antenna (with 3D T-R separation distances ranging from 61 m to 930 m) and 15 TX-RX location combinations were measured using the widebeam RX antenna (with 3D T-R separation distances between 70 m and 728 m) for the UMa scenarios, where for each TX-RX location combination, PDPs for several TX and RX antenna azimuth and elevation pointing angle combinations were recorded. This paper only involves measurement data with narrowbeam antennas (21 LOS omnidirectional data points, and 12 NLOS ones) since it constitutes the majority of the measured data, and defers widebeam studies to future work.

\subsection{UMa Measurements at Aalborg University}
Further, UMa propagation measurements have been performed in Vestby, Aalborg, Denmark, in the 2 GHz, 10 GHz, 18 GHz, and 28 GHz frequency bands in March 2015. Vestby represents a typical medium-sized European city with regular building height and street width, which is approximately 17 m (5 floors) and 20 m, respectively. There were six TX locations, with a height of 15, 20, or 25 m.  A narrowband continuous wave (CW) signal was transmitted at the frequencies of interest, i.e. 10, 18 and 28 GHz, and another CW signal at 2 GHz was always transmitted in parallel and served as a reference. The RX was mounted on a van, driving at a speed of 20 km/h within the experimental area. The driving routes were chosen so that they were confined within the HPBW of the TX antennas. The received signal strength and GPS location were recorded at a rate of 20 samples/s using the R\&S TSMW Universal Radio Network Analyzer for the calculation of path loss and T-R separation distances.  The data points were visually classified into LOS and NLOS conditions based on Google Maps.

\section{Line-Of-Sight Probability in UMa Scenarios}
In this section, the LOS probability model is investigated using only the AAU data, as the UT-Austin data set is too sparse to help the LOS modeling. As mentioned in Section~\ref{Measurements}, the data points were visually classified into LOS and NLOS conditions based on Google Maps. For this study we consider four different models to determine the LOS probabilities using the measured data from AAU. The first is the LOS probability model from the 3GPP 3D channel model in the UMa scenario for a user equipment (UE) height of 1.5 m\cite{3GPP_LTE} which is given as
\begin{equation}\label{3gppLOS_UMa}
p(d)=min\left(\frac{18}{d},1\right)\left(1-e^{-\frac{d}{63}}\right)+e^{-\frac{d}{63}}
\end{equation}

\noindent where $d$ is the distance in $m$. The second model, the 3GPP $d_1/d_2$ model, is similar to Eq.~\eqref{3gppLOS_UMa}:
\begin{equation}\label{3gppLOS}
p(d)=min\left(\frac{d_1}{d},1\right)\left(1-e^{-\frac{d}{d_2}}\right)+e^{-\frac{d}{d_2}}
\end{equation}

\noindent where $d_1$ and $d_2$ are parameters to be optimized to fit the data. It should be noted that the difference between~\eqref{3gppLOS_UMa} and~\eqref{3gppLOS} is that for the UMa scenario, 3GPP has already defined $d_1$ as 18 m and $d_2$ as 63 m, but those values are intended for a base height of 25 m whereas our TX height is 20 m or 25 m. However, it is still instructive to compare the current 3GPP UMa  model to the AAU data. 

The third model is the one proposed by New York University (NYU) in\cite{MKS:WCL15} which is basically the 3GPP $d_1/d_2$ model in~\eqref{3gppLOS} but with a squared term for the LOS probability:
\begin{equation}\label{sqLOS}
p(d)=\left(min\left(\frac{d_1}{d},1\right)\left(1-e^{-\frac{d}{d_2}}\right)+e^{-\frac{d}{d_2}}\right)^2
\end{equation}

\noindent Note that \cite{MKS:WCL15} showed that the squaring gives a better fit to the LOS probability at mmWave frequencies by using a much higher spatial resolution for determining the LOS in a physical database as compared to the original 3GPP model of\cite{3GPP_LTE} for the environments studied which were closer to UMi type environments. 

The final model considered is given by an inverse exponential\cite{Rod13} as
\begin{equation}\label{InvExp}
p(d)=\frac{1}{1+e^{d_1(d-d_2)}}
\end{equation}

\begin{figure}
\centering
 \includegraphics[width=0.45\textwidth]{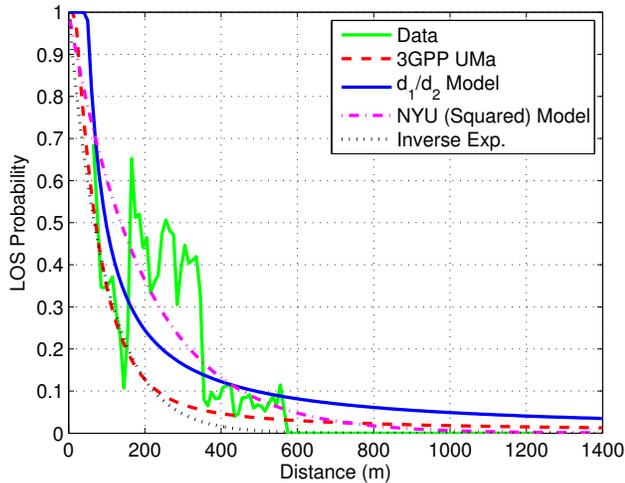}
    \caption{LOS probability for the Aalborg data set plus the four models.}
    \label{fig:LOS}
\end{figure}

For all models we found $d_1$ and $d_2$ that best fit the data in a MMSE sense. In order to smooth the LOS probability for the measured data, a LOS probability versus distance was found for each distance by computing a LOS probability at that given distance using all points within +/-5 m of that distance. Next, the MMSE fitting was done for all distance locations in the data curve (see Fig.~\ref{fig:LOS}), The MMSE fitting for the different models is summarized in Table~\ref{tbl:LOS} and the resulting LOS probabilities are shown in Fig.~\ref{fig:LOS}. As can be seen in Table~\ref{tbl:LOS} and Fig.~\ref{fig:LOS}, the inverse exponential model given in \eqref{InvExp} produced the best fit in terms of the mean square error (MSE). The current 3GPP model for UMa in~\eqref{3gppLOS_UMa} predicts the steepest drop off in the LOS probability within 200 m, where the likelihood of LOS appears greatest from the measured data. The $d_1/d_2$ model in~\eqref{3gppLOS} predicts LOS out to beyond 1 km (as shown by its tail), which is clearly not supported by the measured data. It looks like the NYU model fits the data best except for the void at around 150 m. More data are needed to see if the original 3GPP LOS probability model works. 

\begin{table}
%\captionsetup{width=\textwidth}
\renewcommand{\arraystretch}{1.1}
\begin{center}
\caption{Parameters for the LOS Probability Models Using Aalborg Measurements.}~\label{tbl:LOS}
\begin{tabular}{|c|c|c|c|}
\hline 
 & $d_1$ (m) & $d_2$ (m) & MSE \\ \hline \hline \cline{1-4}
 3GPP UMa & 18 & 63 & 0.0204 \\ \cline{1-4}
 3GPP $d_1/d_2$ & 49 & 1 & 0.0135 \\ \cline{1-4}
 NYU (Squared) & 0 & 395 & 0.0103 \\ \cline{1-4}
 Inv. Exp. & 0.0054 & 97 & 0.0076 \\ \cline{1-4}
\end{tabular}
\end{center}
\end{table} 

\section{Single-Slope Alpha-Beta-Gamma and Close-In Reference Distance Path Loss Models}
Alpha-beta-gamma (ABG) and close-in (CI) free space reference distance models are two candidate multi-frequency large-scale path loss models for 5G cellular communications\cite{MacCartney15_2}. The equation for the ABG model is given by~\eqref{ABG}:
\begin{equation}\label{ABG}
\PL^{\ABG}(f,d)[dB]=10\alpha \log_{10}(\frac{d}{1~m})+\beta+10\gamma \log_{10}(\frac{f}{1~GHz})
\end{equation}

\noindent where $\PL^{\ABG}(f,d)$ denotes the mean path loss in dB over frequency and distance, $\alpha$ and $\gamma$ are coefficients showing the dependence of path loss on distance and frequency, respectively, $\beta$ is the optimized offset in path loss, $f$ is the carrier frequency in GHz, $d$ is the 3D T-R separation distance in meters. The coefficients $\alpha$, $\beta$, and $\gamma$ are obtained through the MMSE method by minimizing the shadow fading standard deviation. 

The equation for the CI model is given by~\eqref{CI}:
\begin{equation}\label{CI}
\PL^{\CI}(f,d)[dB]=\FSPL(f, 1~m)[dB]+10n\log_{10}(\frac{d}{1~m})
\end{equation}

\noindent where $\PL^{\CI}(f,d)$ is the mean path loss in dB over frequency and distance, $n$ represents the path loss exponent (PLE), $d$ is the 3D T-R separation distance, $\FSPL(f,1~m)$ denotes the free space path loss in dB at a T-R separation distance of 1 m at the carrier frequency $f$:
\begin{equation}\label{FSPL}
\FSPL(f,1~m)[dB]=20\log_{10}(\frac{4\pi f}{c})
\end{equation}

\noindent where $c$ is the speed of light. Note that the CI model inherently has an intrinsic frequency dependency of path loss already embedded in it with the 1 m free space path loss value, and it has only one parameter (PLE), as opposed to three parameters in the ABG model ($\alpha, \beta$, and $\gamma$).

While the ABG model offers some physical basis in the $\alpha$ term, being based on a 1 m reference distance, it departs from physics when introducing both an offset $\beta$ (which is basically a floating offset that is not physically based), and a frequency weighting term $\gamma$ which has no proven physical basis for outdoor channels --- in fact, recent work shows the PLE in outdoor mmWave channels to have little frequency dependence\cite{Rap15:TCOM}, whereas indoor channels have noticeable frequency dependence of path loss beyond the first meter\cite{MacCartney15_2}. It is noteworthy that the ABG model is identical to the CI model if we equate $\alpha$ in the ABG model in~\eqref{ABG} with the PLE $n$ in the CI model in~\eqref{CI}, $\gamma$ in~\eqref{ABG} with the free space PLE of 2, and $\beta$ in~\eqref{ABG} with $20\log_{10}(4\pi/c)$  in~\eqref{FSPL}. 

The CI model is based on fundamental principles of wireless propagation, dating back to Friis and Bullington, where the PLE offers insight into path loss based on the environment, having a value of 2 in free space as shown by Friis and a value of 4 for the asymptotic two-ray ground bounce propagation model\cite{Rappaport:Wireless2nd}. Previous
UHF (Ultra-High Frequency)/microwave models used a close-in reference distance of 1 km or 100 m since base station towers were tall without any nearby obstructions and inter-site distances were on the order of many kilometers for those frequency bands\cite{Rappaport:Wireless2nd,Hata:TVT80}. We use $d_0$ = 1 m in mmWave path loss models since base stations will be shorter or mounted indoors, and closer to obstructions~\cite{Rap13:Access,Rap15:TCOM}. The CI 1 m reference distance is a suggested standard that ties the true transmitted power or path loss to a convenient close-in distance of 1 m, as suggested in\cite{Rap15:TCOM}. Standardizing to a reference distance of 1 m makes comparisons of measurements and models simpler, and provides a standard definition for the PLE, while enabling intuition and rapid computation of path loss without a calculator.

\begin{figure}
\centering
 \includegraphics[width=0.45\textwidth]{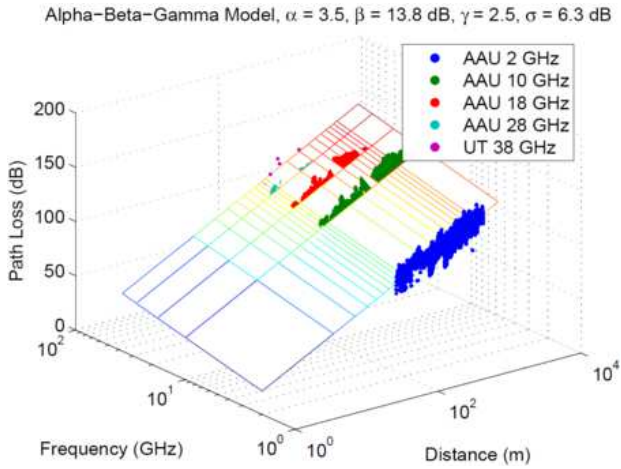}
    \caption{Alpha-beta-gamma path loss model in the UMa scenario across different frequencies and distances in NLOS environments.}
    \label{fig:UMa_NLOS}
\end{figure}

\begin{figure}
\centering
 \includegraphics[width=0.45\textwidth]{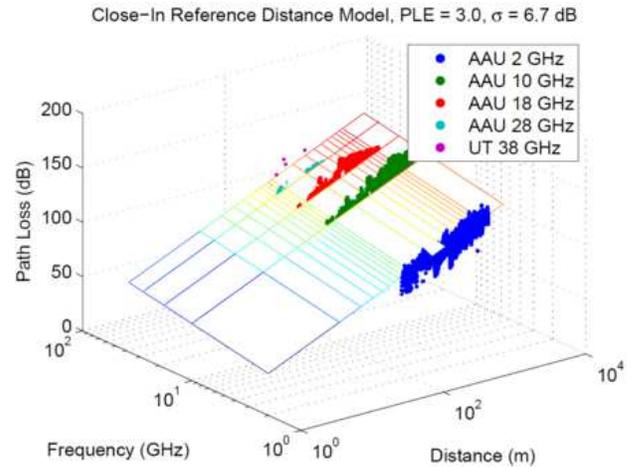}
    \caption{Close-in free space reference distance path loss model in the UMa scenario across different frequencies and distances in NLOS environments.}
    \label{fig:UMa_NLOS_CI}
\end{figure}

Using the two path loss models described above, and the measurement data from UT Austin and AAU, we computed the path loss parameters in the two models. Figs.~\ref{fig:UMa_NLOS} and~\ref{fig:UMa_NLOS_CI} show the ABG and CI models in the UMa scenario in NLOS environments across the five frequencies, respectively. Table~\ref{tbl:ABG_CI} summarizes the path loss parameters in the ABG and CI models for the UMa scenario in both LOS and NLOS environments. As shown by Table~\ref{tbl:ABG_CI}, although the CI model yields slightly higher (by up to 0.4 dB) shadow fading standard deviation than the ABG model, this difference is not significant and is an order of magnitude lower than the actual shadow fading standard deviation in both of the models. This suggests the single-parameter physics-based CI model is suitable for modeling path loss in UMa mmWave channels.

\begin{table*}
%\captionsetup{width=\textwidth}
\renewcommand{\arraystretch}{1.1}
\begin{center}
\caption{Path loss parameters in the CI and ABG models for the UMa scenario.}~\label{tbl:ABG_CI}
\begin{tabu}{|c||c|c|c|c|[1.5pt]c|c|[1.5pt]|[1.5pt]c|c|c|c|[1.5pt]c|c|}
\hline 
\multirow{3}{*}{Scenario} & \multicolumn{6}{c|[1.5pt]|[1.5pt]}{LOS} & \multicolumn{6}{c|}{NLOS}\\ \cline{2-13}
   & \multicolumn{4}{c|[1.5pt]}{ABG Model} & \multicolumn{2}{c|[1.5pt]|[1.5pt]}{CI Model} & \multicolumn{4}{c|[1.5pt]}{ABG Model} & \multicolumn{2}{c|}{CI Model} \\ \cline{2-13}
 & $\alpha$ & $\beta$ (dB) & $\gamma$ & $\sigma$ (dB) & PLE & $\sigma$ (dB) & $\alpha$ & $\beta$ (dB) & $\gamma$ & $\sigma$ (dB)  & PLE & $\sigma$ (dB) \\ \cline{1-13}
UMa & 2.1 & 31.7 & 2.0 & 3.9 & 2.1 & 3.9 & 3.5 & 13.8 & 2.5 & 6.3 & 3.0 & 6.7 \\ \cline{1-13}
%UMi SC & 2.5 & 17.5 & 2.3 & 6.3 & 1.9 & 4.9 & 3.3 & 21.0 & 2.0 & 10.4 & 2.8 & 10.1 \\ \cline{1-13}
%UMi OS & 2.6 & 23.5 & 1.5 & 3.8 & 1.9 & 4.7 & 4.6 & 1.1 & 1.7 & 7.6 & 2.8 & 8.3 \\ \cline{1-13}
\end{tabu}
\end{center}
\end{table*}

\section{Single-Slope and Dual-Slope Path Loss Models}
Four types of large-scale path loss models are studied in this section using measured data: the single-slope CI model, the single-slope FI model, the dual-slope CI model, and the dual-slope FI model. 
%The single-slope CI and FI path loss models are canonic models, with only the CI model being based on propagation physics\cite{Rap15:TCOM}, that have been widely adopted in previous literature on wireless channel models\cite{Rappaport:Wireless2nd,RapGut13,Rap13:Access,Rap15,Mac13,Mac14:PIMRC}.

Dual-slope path loss models for both the CI and FI models are investigated to provide comprehensive analyses. The dual-slope path loss equation for the CI model is as follows
\begin{equation}
\PL^{\CI}_{Dual}(d)=\left\{\begin{array}{lcl}
\FSPL(1m)+10n_1\log_{10}(d) &\mbox{for} & d\leq d_{th} \\ \FSPL(1m)+10n_1\log_{10}(d_{th})\\+10n_2\log_{10}(d/d_{th})& \mbox{for}
& d>d_{th} 
\end{array}\right.
\end{equation}

\noindent where $\PL$ denotes the mean path loss in dB as a function of the 3D distance $d$, $\FSPL$ represents free space path loss in dB, $d_{th}$ is the threshold distance (also called the breakpoint\cite{Bla92,Feu94}) in meters, $n_1$ is the PLE for distances smaller than $d_{th}$, and $n_2$ is the slope of the average path loss for distances larger than $d_{th}$. The dual-slope equation for the FI model is given below
\begin{equation}
\PL^{\FI}_{Dual}(d)=\left\{\begin{array}{lcl}
\alpha_1+10\beta_1\log_{10}(d) &\mbox{for} & d\leq d_{th} \\ \alpha_1+10\beta_1\log_{10}(d_{th})\\+10\beta_2\log_{10}(d/d_{th})& \mbox{for}
& d>d_{th} 
\end{array}\right.
\end{equation}

\noindent where $\alpha_1$ denotes the floating intercept, $\beta_1$ and $\beta_2$ are the two slopes for different distance ranges\cite{Mac13,Rap15:TCOM,Soo15:JSTSP}. Both of the dual-slope FI and CI models are continuous functions of distance. The criterion for finding the $d_{th}$ is to minimize the global standard deviation of the shadow fading, i.e., to iteratively set all the possible distances as the breakpoint (from the smallest to largest measured distances in 1 m increment), calculate the two slopes, check the resultant RMS error versus distance, and find the distance corresponding to the minimum RMS error.

\begin{figure}
\centering
 \includegraphics[width=0.45\textwidth]{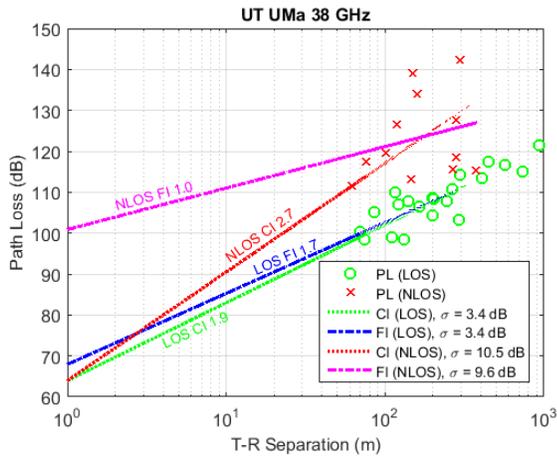}
    \caption{Single-slope CI and FI omnidirectional path loss models for the 38 GHz UMa scenario (data from\cite{RapGut13,Rap15:TCOM}). $\sigma$ denotes the standard deviation of shadow fading.}
    \label{fig:UT_38GHz}
\end{figure}

\begin{figure}
\centering
 \includegraphics[width=0.45\textwidth]{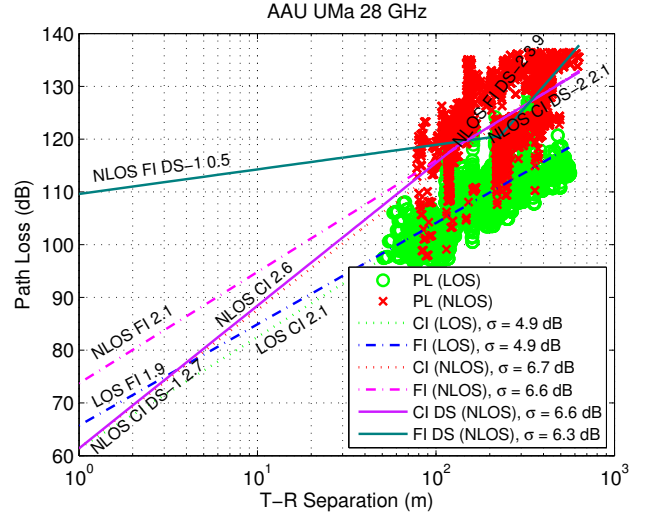}
    \caption{Single-slope and dual-slope CI and FI omnidirectional path loss models for the 28 GHz UMa scenario. $\sigma$ denotes the standard deviation of shadow fading.}
    \label{fig:AAU_18GHz}
\end{figure}

%\begin{figure}
%\centering
% \includegraphics[width=3.7in]{Aalborg_UMa_10GHz.png}
   % \caption{Single-slope and dual-slope CI and FI omnidirectional path loss models for the 10 GHz UMa scenario. $\sigma$ denotes the standard deviation of shadow fading.}
   % \label{fig:AAU_10GHz}
%\end{figure}

%\begin{figure}
%\centering
% \includegraphics[width=3.7in]{Aalborg_UMa_2GHz.png}
   % \caption{Single-slope and dual-slope CI and FI omnidirectional path loss models for the 2 GHz UMa scenario. $\sigma$ denotes the standard deviation of shadow fading.}
   % \label{fig:AAU_2GHz}
%\end{figure}

\begin{table*}
%\captionsetup{width=\textwidth}
\renewcommand{\arraystretch}{1.1}
\begin{center}
\caption{Large-scale parameters in path loss and shadow fading models. dth represents the threshold distance. "Dual" refers to the dual-slope path loss model. Note that the negative slopes (marked with $^*$) are not usable due to a low number of samples.}~\label{tbl:DS}
\begin{tabu}{|c|c||c|c|[1.7pt]c|c|[1.7pt]c|c|[1.7pt]c|c|[1.7pt]c|c|}
\hline 
 \multicolumn{2}{|c||}{} &\multicolumn{10}{c|}{\textbf{UMa}} \\ \cline{3-12}
 \multicolumn{2}{|c||}{}& \multicolumn{2}{c|[1.7pt]}{\textbf{UT 38 GHz\cite{RapGut13,Rap15:TCOM}}} & \multicolumn{2}{c|[1.7pt]}{\textbf{AAU 28 GHz}}& \multicolumn{2}{c|[1.7pt]}{\textbf{AAU 18 GHz}} & \multicolumn{2}{c|[1.7pt]}{\textbf{AAU 10 GHz}} & \multicolumn{2}{c|}{\textbf{AAU 2 GHz}}\\ \cline{3-12}
 \multicolumn{2}{|c||}{} & LOS & NLOS & LOS & NLOS & LOS & NLOS & LOS & NLOS & LOS & NLOS \\ \tabucline[1.7pt red off 0pt]{1-12}
\multirow{2}{*}{\textbf{CI}} & PLE & 1.9 & 2.7 & 2.1 & 2.6 & 2.1 & 3.1 & 2.2 & 3.2 & 2.1 & 2.9 \\ \cline{2-12}
 & \text{$\sigma$ [dB]} & 3.4 & 10.5 & 4.9 & 6.7 & 4.6 & 5.8 & 5.5 & 7.2 & 3.3 & 7.0 \\ \cline{1-12}
\multirow{4}{*}{\textbf{CI Dual}} & $d_{th}$ [m] & \multirow{4}{*}{N/A} & 205 & \multirow{4}{*}{N/A} & 129 & \multirow{4}{*}{N/A} & 368 &\multirow{4}{*}{N/A} & 698 & \multirow{4}{*}{N/A} & 381 \\ \cline{2-2} \cline{4-4} \cline{6-6} \cline{8-8} \cline{10-10} \cline{12-12}
 & $n_1$ & & 2.9 & & 2.7 & & 3.1 & & 3.2 & & 2.9 \\ \cline{2-2} \cline{4-4} \cline{6-6} \cline{8-8} \cline{10-10} \cline{12-12}
 & $n_2$ & & -4.4$^*$ & & 2.1 & & 3.3 & & 0.9 & & 3.9 \\ \cline{2-2} \cline{4-4} \cline{6-6} \cline{8-8} \cline{10-10} \cline{12-12}
 & $\sigma$ [dB] & & 8.6 & & 6.6 & & 5.8 & & 7.0 & & 6.9 \\ \tabucline[1.7pt red off 0pt]{1-12}
\multirow{3}{*}{\textbf{FI}} & $\alpha [dB]$ & 67.9 & 100.9 & 65.7 & 73.7 & 57.4 & 47.6 & 58.5 & 54.2 & 39.9 & 27.9 \\ \cline{2-12}
 & $\beta$ & 1.7 & 1.0 & 1.9 & 2.1 & 2.1 & 3.4 & 1.9 & 3.1 & 2.0 & 3.3 \\ \cline{2-12}
 & $\sigma$ [dB] & 3.4 &  9.6 & 4.9 & 6.6 & 4.6 & 5.8 & 5.5 & 7.2 & 3.3 & 7.0 \\ \cline{1-12}
\multirow{5}{*}{\textbf{FI Dual}} & $d_{th}$ [m] & \multirow{4}{*}{N/A} & 184 & \multirow{4}{*}{N/A} & 229 & \multirow{4}{*}{N/A} & 134 &\multirow{4}{*}{N/A} & 634 & \multirow{4}{*}{N/A} & 375 \\ \cline{2-2} \cline{4-4} \cline{6-6} \cline{8-8} \cline{10-10} \cline{12-12}
 & $\alpha_1$ [dB] & & 32.0 & & 109.6 & & -49.2 & & 34.0 & & 42.1 \\ \cline{2-2} \cline{4-4} \cline{6-6} \cline{8-8} \cline{10-10} \cline{12-12}
 & $\beta_1$ & & 4.5 & & 0.5 & & 8.1 & & 3.9 & & 2.7 \\ \cline{2-2} \cline{4-4} \cline{6-6} \cline{8-8} \cline{10-10} \cline{12-12}
 & $\beta_2$ & & -4.4$^*$ & & 3.9 & & 3.1 & & 0.8 & & 3.9 \\ \cline{2-2} \cline{4-4} \cline{6-6} \cline{8-8} \cline{10-10} \cline{12-12}
& $\sigma$ [dB] & & 8.4 & & 6.3 & & 5.6 & & 6.9 & & 6.9 \\ \cline{1-12}
\end{tabu}
\end{center}
\end{table*} 

Using the methodology described above, we processed the path loss data from the UT Austin and AAU measurements for both single- and dual-slope models. Figs.~\ref{fig:UT_38GHz} and~\ref{fig:AAU_18GHz} illustrate the scatter plots of path loss data for the four models at 38 GHz measured on the campus of UT Austin, and at 28 GHz measured at AAU, respectively. As shown by Figs.~\ref{fig:UT_38GHz} and~\ref{fig:AAU_18GHz}, for either 38 GHz or 28 GHz in either the LOS or NLOS environment, the scattered path loss data points do not exhibit obvious dual-slope trends, i.e., there is no visible breakpoint such that the changing rate of path loss versus distance is significantly different before and after the breakpoint. Table~\ref{tbl:DS} lists the large-scale parameters for path loss and shadow fading in the four path loss models. As shown by Table~\ref{tbl:DS}, in LOS environments, the PLE ranges from 1.9 to 2.2 using the single-slope CI model, which matches well with the free space propagation with a PLE of 2. The slope in the single-slope FI model lies between 1.7 and 2.1 for LOS environments. Note that the standard deviations of shadow fading are very small for both the single-slope CI and FI models in LOS environments. The NLOS CI PLE is between 2.6 and 3.2 for frequencies ranging from 2 GHz to 38 GHz for UMa scenarios, which is comparable to the PLEs observed in current cellular communication systems. The slope for NLOS FI models varies from 1.0 to 3.4, showing that the FI model is much more sensitive to the geometrical environment, measured distances, and the number of data samples, when compared to the CI model. Furthermore, the CI model exhibits consistent PLEs across frequencies and environments, hence is preferable over the FI model since the PLE can be a single value for all frequencies as long as oxygen absorption is not a factor\cite{Rap15:TCOM}. 

For a fixed frequency, comparing the standard deviations of shadow fading between the single-slope model and the corresponding dual-slope model, we can see that the dual-slope model does reduce the RMS error in all the investigated cases. For instance, considering the 28 GHz UMa scenario, the standard deviation of the shadow fading is reduced by 0.1 dB from 6.7 dB in the single-slope CI model to 6.6 dB in the dual-slope CI model. Although the standard deviation of shadow fading is slightly smaller using the FI model compared to the CI model, the difference is within 1 dB for both single-slope and dual-slope cases, which is negligible given the typical standard deviation value of 6 dB to 10 dB for shadow fading. Therefore, the dual-slope CI model is preferable to its FI counterpart in terms of its physical basis and consistency when comparing path loss values across different frequencies, measurement campaigns, and research groups throughout the world, as suggested in\cite{Rap15:TCOM}. 

It is noteworthy that although the dual-slope model can improve the RMS error, the improvement is no more than 0.3 dB in most cases\footnote{The 38 GHz data did show a 1.9 dB improvement with the dual-slope CI model but it also produced a negative second slope.}. Additionally, the threshold distance varies substantially across frequencies, revealing the frequency-dependence feature of the threshold distance. Given the above characteristics, the dual-slope model seems unnecessary and unduly complex for UMa scenarios at cmWave and mmWave frequencies, at least over the distance range studied, while it could be well needed for larger distances.

\section{Distance-Dependent Shadow Fading Models}
\begin{figure}
\centering
 \includegraphics[width=0.45\textwidth]{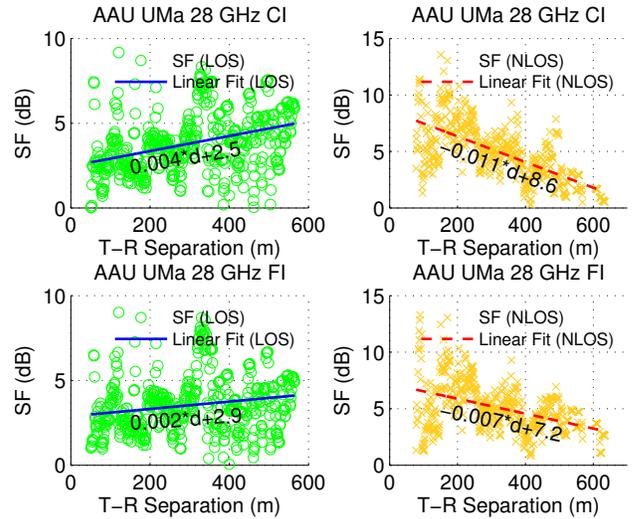}
    \caption{Standard deviation of shadow fading as a function of T-R separation distance using both the CI and FI path loss models for the 28 GHz UMa scenario.}
    \label{fig:28GHz_SF}
\end{figure}

%\begin{figure}
%\centering
% \includegraphics[width=3.7in]{Aalborg_UMa_18GHz_SF.png}
   % \caption{Standard deviation of shadow fading as a function of T-R separation distance using both the CI and FI path loss models for the 18 GHz UMa scenario.}
   % \label{fig:AAU_18GHz_SF}
%\end{figure}

%\begin{figure}
%\centering
 %\includegraphics[width=3.7in]{Aalborg_UMa_10GHz_SF.png}
   %\caption{Standard deviation of shadow fading as a function of T-R separation distance using both the CI and FI path loss models for the 10 GHz UMa scenario.}
    %\label{fig:AAU_10GHz_SF}
%\end{figure}

%\begin{figure}
%\centering
% \includegraphics[width=3.7in]{Aalborg_UMa_2GHz_SF.png}
   % \caption{Standard deviation of shadow fading as a function of T-R separation distance using both the CI and FI path loss models for the 2 GHz UMa scenario.}
    %\label{fig:AAU_2GHz_SF}
%\end{figure}

\begin{table*}
%\captionsetup{width=\textwidth}
\renewcommand{\arraystretch}{1.1}
\begin{center}
\caption{Parameters in UMa shadow fading models with respect to the T-R separation distance.}~\label{tbl:SF}
\begin{tabu}{|c|c||c|c|[1.7pt]c|c|[1.7pt]c|c|[1.7pt]c|c|[1.7pt]c|c|}
\hline 
 \multicolumn{2}{|c||}{} &\multicolumn{10}{c|}{\textbf{UMa}} \\ \cline{3-12}
 \multicolumn{2}{|c||}{}& \multicolumn{2}{c|[1.7pt]}{\textbf{UT 38 GHz}} & \multicolumn{2}{c|[1.7pt]}{\textbf{AAU 28 GHz}} & \multicolumn{2}{c|[1.7pt]}{\textbf{AAU 18 GHz}} & \multicolumn{2}{c|[1.7pt]}{\textbf{AAU 10 GHz}} & \multicolumn{2}{c|}{\textbf{AAU 2 GHz}}\\ \cline{3-12}
 \multicolumn{2}{|c||}{} & LOS & NLOS& LOS & NLOS & LOS & NLOS & LOS & NLOS & LOS & NLOS \\ \tabucline[1.7pt red off 0pt]{1-12}
\multirow{2}{*}{\textbf{CI}} & A & -0.002 & 0.03 & 0.004 & -0.011 & 0.002 & -0.005 & 0.008 & -0.003 & 0.002 & -0.003 \\ \cline{2-12}
 & B [dB] & 3.3 & 2.6 & 2.5 & 8.6 & 2.9 & 6.2 & 1.9 & 7.3 & 2.1 & 6.8 \\ \tabucline[1.7pt red off 0pt]{1-12}
\multirow{2}{*}{\textbf{FI}} & A & -0.001 & 0.02 & 0.002 & -0.007 & 0.002 & -0.004 & 0.005 &  -0.003 & 0.002 & -0.003 \\ \cline{2-12}
 & B [dB] & 3.1 & 4.8 & 2.9 & 7.2 & 2.8 & 5.9 & 2.9 & 7.5 & 2.1 & 6.9 \\ \cline{1-12}
\end{tabu}
\end{center}
\end{table*} 

In this section, the magnitude of shadow fading is analyzed and modeled as a function of the 3D T-R separation distance from the AAU and UT Austin measurements, using both the CI and FI single-slope path loss models. The relationship between the shadow fading magnitude and the T-R separation distance is modeled as follows
\begin{equation}\label{std}
\SF~[\dB]=A*d+B
\end{equation}

\noindent where $\SF$ represents the shadow fading magnitude, $A$ reflects the changing rate of $\SF$ over distance, $d$ is the 3D T-R separation distance in meters, and $B$ is the intercept determined by MMSE linear fit on $\SF$. 

 Fig.~\ref{fig:28GHz_SF} displays the scatter plots and fitted linear models of the shadow fading magnitude over the 3D T-R separation distance for 28 GHz data measured in the UMa scenario, where the shadow fading magnitude is obtained by averaging the shadow fading magnitudes over a distance bin width of 1 m. The parameters for modeling the relationship between shadow fading magnitude and the distance at 2 GHz, 10 GHz, 18 GHz, 28 GHz, and 38 GHz are summarized in Table~\ref{tbl:SF}. The fitted linear model is obtained through MMSE linear fit on the local RMS error at each individual distance bin. For the 28 GHz UMa scenario, the shadow fading magnitude in LOS environments slightly increases with the T-R separation distance for both of the CI and FI path loss models; in contrast, the NLOS shadow fading magnitude decreases with distance. Similar phenomena are observed at 2 GHz, 10 GHz, and 18 GHz. This observation may be due to limited measurement range, for which larger distances have fewer detectable measurements, causing a clustering of detected energy. For the UT data, the opposite is observed, where the LOS shadow fading magnitude exhibits a slight decreasing trend over distance while the NLOS shadow fading magnitude increases with distance. However, the number of data points in the 38 GHz measurement set is relatively small. Based on the current available data, it seems that the shadow fading magnitude increases with distance in LOS environments and decreases with distance in NLOS environments, but the decreasing shadow fading magnitude may be caused by measurement range limitations. Therefore, further study is encouraged to gain more insight on the issue.

\section{Conclusion}
In this paper, we presented the LOS probability, multi-frequency ABG and CI omnidirectional path loss models, single- and dual-slope CI and FI omnidirectional path loss models, and distance-dependent shadow fading in the UMa scenario, using the data at 2 GHz, 10 GHz, 18 GHz, and 28 GHz measured in Aalborg, Denmark, and at 38 GHz measured in UT Austin, USA. The LOS probability should be explored further, since cells will likely be smaller in mmWave systems, where more spatial resolution will be needed in such models. The ABG and CI models are both potential omnidirectional path loss models to be considered for the UMa scenario, and it should be noted that the ABG model is similar to the FI model in that offsets are used, while the CI model has a physical tie to transmitted power and has a frequency-dependent path loss factor in the first meter, causing PLEs to be much more similar over wide ranges of frequency, with virtually identical shadowing standard deviation compared to the ABG or FI model. The dual-slope omnidirectional path loss model was able to slightly reduce the RMS error of path loss versus distance in comparison with its single-slope counterpart, but by no more than 0.3 dB in most cases, thus it is likely not worth using given the extra computational complexity. Regarding shadow fading in the UMa scenario, the magnitude of shadow fading seems to increase with distance for LOS while decreasing with distance for NLOS, as suggested by the measured AAU data, but this may be because of the limited measurement range where larger distances have fewer detectable measurements. Further measurements, especially at larger distances, are encouraged to improve the understanding of the UMa scenario at cmWave and mmWave frequencies.

%\section*{Acknowledgment}

\ifCLASSOPTIONcaptionsoff
  \newpage
\fi

\bibliographystyle{IEEEtran}
\bibliography{bibliography}

\end{document}